\documentclass[doublecol, figures]{epl2} % for 2 columns style with line numbers
\usepackage[english]{babel} 			% The last language is the active one
\usepackage[utf8]{inputenc} 			% Text encoding UTF8
\usepackage[T1]{fontenc}				% Output encoding
\usepackage{lmodern}					% Enhanced versions of the CM fonts
\usepackage{amsmath} 					% For maths
\usepackage{amssymb} 					% For maths
\usepackage{bbold}   					% Enable the \textbb command
\newcommand{\mean}[1]{\langle#1\rangle}						% Mean value
\newcommand{\td}[2]{\frac{\mathrm{d} #1}{\mathrm{d} #2}}	% Total derivative
\newcommand{\pd}[2]{\frac{\partial #1}{\partial #2}}		% Partial derivative
\newcommand{\dd}{\mathrm{d}}      						% Differential d for integrals
\newcommand{\D}{{\cal D}}      							% D for  path integrals
\newcommand{\eps}{\varepsilon}								% Epsilon
\newcommand{\norm}[1]{\left|#1\right|}                      % Norm
\newcommand{\rhob}{\bar{\rho}}								% rho bar
\newcommand{\vb}{\bar{v}}									% v bar

\title{Large-scale fluctuations of the largest Lyapunov exponent in diffusive systems}

\author{T. Laffargue\inst{1} \and P. Sollich\inst{2} \and J. Tailleur\inst{1} \and F. van Wijland\inst{1}}
\shortauthor{T. Laffargue \etal}

\institute{                    
  \inst{1} Laboratoire Matière et Systèmes Complexes, UMR 7057 CNRS/P7, 
			 Université Paris Diderot, 10 rue Alice Domon et Léonie Duquet, 75205 Paris Cedex 13, France\\
  \inst{2} Department of Mathematics, King's College London, Strand, London WC2R 2LS, United Kingdom
}

\abstract{
We present a general formalism for computing the largest Lyapunov exponent and its fluctuations in spatially extended systems described by diffusive fluctuating hydrodynamics, thus extending the concepts of dynamical system theory to a broad range of non-equilibrium systems. Our analytical results compare favourably with simulations of a lattice model of heat conduction. We further show how the computation of the Lyapunov exponent for the Symmetric Simple Exclusion Process relates to damage spreading and to a two-species pair annihilation process, for which our formalism yields new finite size results.}

\begin{document}
\maketitle

The long history of cross-fertilisation between statistical mechanics and chaos theory has led to the emergence of many subfields where they are deeply intertwined~\cite{Dorfman1999}, from the foundation of statistical physics~\cite{Penrose1979, Gaspard1998a, Dettmann1999, Grassberger1999}, to the discovery of the fluctuation theorems~\cite{Evans1993, Gallavotti1995} and the study of random dynamical
systems~\cite{Benzi1984, Arnold1986, Graham1988, Paladin1995}. Paramount here is the idea that studying fluctuations of dynamical observables could allow one to extend statistical mechanics from configuration space into trajectory space. Indeed, while the usual statistical mechanics tells us about the phases of a given system, it remains silent about their dynamical nature. Working in trajectory space allows one to answer this question in an elegant way.

Over the past ten years, a number of methods have been found for studying fluctuations directly in trajectory space~\cite{Derrida2004, Bodineau2004, Bertini2005, Bertini2005a, Imparato2009, Brunet2010, Lecomte2010, Hurtado2011, Meerson2014, Garrahan2007, Turci2011, Lecomte2012, Speck2012} whence revealing novel dynamical phase transitions. Perhaps the most salient example is that of glassy systems, whose statics do not differ from their liquid counterpart but whose dynamics display a drastic slowing down. These have recently been studied by classifying trajectories according to their level of dynamic activity~\cite{Merolle2005, Garrahan2007, Hedges2009, Pitard2011}. While the activity is easy to define for lattice-based kinetically constrained models~\cite{Garrahan2007}, quantifying it in realistic physical systems such as molecular glasses has always involved a great deal of arbitrariness: the need to distinguish cooperatively rearranging regions from local rattling leads to \textit{ad hoc} constructions based on \textit{a posteriori} knowledge of the dynamic evolution~\cite{Pitard2011,Speck2012,Fullerton2013}. 

A natural path to circumvent this problem is to rely on more fundamental quantities, such as the Lyapunov exponents (LEs) that form the basis of the thermodynamic formalism of Bowen, Ruelle and Sinai~\cite{Ruelle1978}. Connections between the Lyapunov spectrum and transport coefficients have been investigated in the recent past~\cite{Dorfman1999,Gaspard1998} and suggest that dynamical phase transitions involving the current of some conserved quantity could also be understood in terms of fluctuations of the Lyapunov spectrum. In fact, Lyapunov exponents may well prove to be the unifying concept behind the variety of known dynamical phase transitions. Unfortunately, studying their fluctuations is a notoriously difficult task that, in spite of a large effort from the community, has been carried out mostly in low dimensions~\cite{Bohr1987, Grassberger1988, Beck1993} (with some notable exceptions~\cite{Appert1997, Kuptsov2011, Pazo2013, Tailleur2007, Laffargue2013}). For deterministic systems, computations in high dimensions appear out of reach, beginning with
the difficult task to find their SRB measures, which are crucial to properly define averages and fluctuations.  Fortunately, many systems
of interest effectively have, to an excellent level of approximation, stochastic dynamics. Then ergodic issues are bypassed and fluctuations are easier to access as they correspond to different noise realisations. There are several ways to define LEs for stochastic
dynamics depending on context and goals~\cite{Benzi1984, Arnold1986, Graham1988,Paladin1995} but studying their fluctuations in high dimensions remains very challenging, with few results available~\cite{Tailleur2007, Laffargue2013}.

Of course, real condensed matter systems are spatially extended and endowed with interactions; the study of chaotic properties of high-dimensional systems is thus of great interest~\cite{Takeuchi2013, Yang2009}. When studying collective phenomena, like the glass transition, our interest is not in the individual behavior of single particles but rather in the emergent behavior of the system. In other words, we are interested in collective modes, rather than in microscopic degrees of freedom. Characterising the fluctuations of LEs
of collective modes is thus both an important goal and a difficult task.

In this Letter, we show how this program can actually be carried out analytically for a class of many-body interacting systems, whose
dynamics is described by diffusive fluctuating hydrodynamics. Such a description applies to systems devoid of long-range interactions (for which special precautions must be taken~\cite{Touchette2006}) and, despite being intuitively appealing, can be mathematically challenging to establish~\cite{Caglioti1996}. For the sake of concreteness, we first introduce a paradigmatic example of such models: the
Kipnis-Marchioro-Presutti (KMP) model of heat conduction~\cite{Kipnis1982}. Then, we show how the Macroscopic Fluctuation Theory (MFT)~\cite{Spohn1983, Bertini2001, Bertini2002, Bertini2005, Bertini2005a, Tailleur2008, Imparato2009, Derrida2009a, Lecomte2010, Meerson2014, Bouchet2014}, which has proven successful in the study of current or activity fluctuations, can be extended to calculate the large-deviation function of the largest LE. We validate our MFT-based analytical results using simulations of a lattice model. Finally, we present how our results on the LEs connect, somewhat unexpectedly, to damage spreading and to a two-species pair annihilation reaction-diffusion process.

The KMP model is a chain of $L$ oscillators\footnote{Periodic boundary conditions are assumed throughout the entire article.}, in which the energy $\eps_i \geqslant 0$ is redistributed stochastically between nearest neighbours at fixed rate $\gamma$ according to: 

\begin{equation*}
	\left( \eps_j, \, \eps_{j+1} \right) \xrightarrow[]{\text{rate }\gamma} \left( p \left(\eps_j + \eps_{j+1}\right), \, (1-p) \left(\eps_j + \eps_{j+1}\right) \right)
\end{equation*}
where for each event $p$ is sampled from a uniform distribution on $[0,1]$. The total energy is conserved in each update, which accounts for this model being one of the simplest for which Fourier's law can be proven analytically~\cite{Kipnis1982}.

To define the LEs, let us consider two copies of the system, $\{\eps_{i}\}$ and $\{\eps'_{i}\}$, which evolve with the same noise realisations. In practice this means taking the same redistribution time (given by an exponential law of parameter $\gamma$) for each bond and the same redistribution parameter $p$ at each activation in the two copies. We can then follow the time evolution of the difference between the two copies, $u_{i} = \eps'_{i} - \eps_{i}$, and define from this the largest (finite-time) LE $\tilde{\lambda}(t)$ as

\begin{equation}
	\tilde{\lambda}(t) \equiv \frac{1}{t} \ln \frac{\norm{\mathbf{u}(t)}}{\norm{\mathbf{u}(0)}}\,, \quad \text{with} \quad  \norm{\mathbf{u}(t)}^2 \equiv {\sum_{i=1}^{L} u_{i}^{2}(t)}\,.
\end{equation}
The LE $\tilde{\lambda}(t)$ -- we omit the adjective ``largest'' below -- tells us how small perturbations are amplified or eliminated by the dynamics. If $\tilde{\lambda}(t) < 0$, the copies of the systems converge towards identical energy profiles. Conversely, if $\tilde{\lambda}(t) > 0$, the difference between the two copies diverges, and a small perturbation on the initial configuration completely changes its subsequent evolution. Since this system is stochastic, generic initial conditions are quickly forgotten and the LE should not depend on them in the large-time limit.

It would be a formidable task to keep track of the $L$ individual stochastic variables. Since we are interested in the macroscopic properties of our model, we adopt a fluctuating hydrodynamics description, which accounts for the stochastic evolution of collective modes in the large $L$ limit. In this approach, space and time are rescaled by the system length and the diffusive relaxation time of a macroscopic fluctuation: $x=i/L$ and $\tau=t/L^2$. The local energy $\eps_{i}(t)$ then turns into a smoothly varying field $\rho(x, \tau)$, which evolves according to a continuity equation

\begin{equation}
	\partial_{\tau} \rho(x, \tau) + \partial_{x} \,j(x, \tau) = 0\,.
	\label{FH}
\end{equation}
The current $j(x,\tau)$ comprises a deterministic contribution arising from Fick's law and a stochastic one accounting for the fluctuations around this typical behaviour:

\begin{equation}
	j(x, \tau) = - D(\rho)\,\partial_{x} \rho - \sqrt{\frac{\sigma(\rho)}{L}}\,\xi(x, \tau)
	\label{current}
\end{equation}
where $\xi(x, \tau)$ is a Gaussian white noise with correlations $\mean{\xi(x, \tau) ~\xi(x', \tau')} = \delta(x-x') ~\delta(\tau - \tau')$. Equation~\eqref{current} shows the benefit of replacing microscopic variables by a continuous stochastic field: the noise vanishes in the large $L$ limit.

This description is generic for conserved quantities in diffusive systems and we can recover different microscopic models by appropriate choice of the $\rho$-dependence of $D$ and $\sigma$. The KMP model with $\gamma=2$ has $D(\rho) = 1$ and $\sigma(\rho) = 2 \rho^{2}$. Equations~\eqref{FH} and~\eqref{current} can also describe the local number of particles in lattice gas models: $D(\rho) = 1$ and $\sigma(\rho) = 2 \rho$ corresponds to free particles performing a symmetric random walk with a unit hopping rate, whereas $D(\rho) = 1$ and $\sigma(\rho) = 2 \rho (1-\rho)$ corresponds to the Symmetric Simple Exclusion Process (SSEP). For the sake of generality, we will keep $D$ and $\sigma$ arbitrary for now.  A small perturbation $u(x,\,\tau)$ of the field $\rho(x,\,\tau)$ evolves according to the linearisation of the continuity equation~\eqref{FH}

\begin{equation*}
  \partial_{\tau} u(x, \, \tau) = \mathbf{A}\, u(x, \, \tau);\:
\mathbf{A} = \pd{^{2}}{x^{2}} D(\rho) + \pd{}{x} \frac{\sigma'(\rho)}{2 \sqrt{L \sigma(\rho)}} \xi
\end{equation*}
where $\xi$ is the same noise as in (\ref{current}) and the differential operator $\pd{}{x}$ applies to everything on its right. Linearising the dynamics amounts to considering two close-by copies of the system $\rho$ and $\rho'$, and to examining the evolution of the difference $u = \rho' - \rho$. In this formalism, the definition of the ``same noise'' is straightforward: we simply take the exact same realisation of $\xi(x, \, \tau)$ for the two copies of our system. The LE $\lambda$ is then defined as

\begin{equation*}
	\lambda(\tau) \equiv \frac{1}{\tau} \ln \frac{\norm{u(\tau)}}{\norm{u(0)}}\,, \; \text{with} \;  \norm{u(\tau)}^2 \equiv {\int_{0}^{1} \dd x \, u^2(x, \tau)}\,.
\end{equation*}

We can now introduce the normalised tangent vector $v(x, \, \tau) = \frac{u(x, \,   \tau)}{\norm{u(\tau)}}$, which evolves according to

\begin{equation*}
	\partial_{\tau} v(x, \, \tau) = \mathbf{A}\, v(x, \, \tau) - v(x, \, \tau) \int_{0}^{1} \dd y \, v(y, \, \tau) \, \mathbf{A} v(y, \, \tau)\,,
\end{equation*} 
to obtain an explicit expression for $\lambda(\tau)$ as~\cite{Laffargue2013}

\begin{equation}
	\lambda(\tau) = \frac{1}{\tau} \int_{0}^{\tau} \dd \tau \, \int_{0}^{1} \dd x \, v(x, \, \tau) \, \mathbf{A} v(x, \, \tau)\,.
\end{equation}

The LE is a fluctuating quantity that depends on the noise realisation. To characterise its fluctuations, it is convenient to introduce the moment-generating function

\begin{equation}
	Z(\alpha, L, \tau) \equiv \mean{e^{\alpha L \tau \lambda(\tau)}}  = \int \dd \lambda ~P(\lambda, L, \tau) ~e^{\alpha L \tau \lambda(\tau)}
\end{equation}
instead of trying to directly calculate the probability distribution $P(\lambda, L, \tau)$. In analogy to the canonical ensemble in equilibrium statistical physics the parameter $\alpha$, which is conjugate to the LE, plays the role of an inverse temperature for chaoticity. Taking $\alpha > 0$ favours trajectories with large LE, i.e.~abnormally chaotic trajectories, whereas $\alpha < 0$ favours trajectories with small LE that are abnormally stable. Our next step is technical: we carry out the evaluation of the partition function $Z(\alpha)$.

Using standard path-integral methods~\cite{Janssen1976, DeDominicis1976, Tailleur2008, Imparato2009, Lecomte2010}, $Z$ can be expressed as

\begin{equation*}
	Z(\alpha, L, \tau) = \int \D \left[ \rho, \rhob, v, \vb\right] e^{- L S[\rho, \rhob, v, \vb]}
\end{equation*}
where the explicit dependence on the noise is replaced by response fields $\rhob$ and $\vb$ and the path integral has to be performed over fields that respect the constraints ${\int_{0}^{1} \dd x \, \rho(x,\tau)}=\rho_0$, with $\rho_0$ the overall density, ${\int_{0}^{1} \dd x \, v(x,\tau)}=0$,  ${\int_{0}^{1} \dd x \, v^{2}(x,\tau)}=1$ and the periodic boundary conditions (for the fields and their first derivatives). The action $S$ reads
\begin{multline}
S = \int_{0}^{\tau} \dd t \int_{0}^{1} \dd x \,\Big[ 
			\rhob \, \partial_{t} \rho  
			+ \vb \, \partial_{t} v 
			+ D \, \partial_{x} \rho \, \partial_{x} \rhob 			 
			\\
			+ \left( \left( I - \alpha \right) v- \vb \right) \partial_{x}^{2} \left( D \, v \right) 
			- \frac{\left( \mathcal{J} + D \, \partial_{x} \rho \right)^{2}}{2 \sigma} 
			\Big]
			\label{eqn:action}
\end{multline}
\begin{align*}
	\text{where} \qquad I &= \int_{0}^{1} \dd y \, v(y, \,t) \, \vb(y, \, t)\\
\text{and}\qquad	\mathcal{J} &= - D \, \partial_{x} \rho + \frac{\sigma'}{2} v \, \partial_{x} \left[ ( I - \alpha) v - \vb \right] - \sigma \, \partial_{x} \rhob
\end{align*}
have been introduced to make $S$ as compact as possible. 

The specific form of the path integral, with system size $L$ factored out in front of the action, gives the gist of the MFT: we may use a saddle-point approximation~\cite{Tailleur2008, Imparato2009, Lecomte2010} to compute $Z$ in the large $L$ limit

\begin{equation}
Z(\alpha, \, L, \, \tau) \approx e^{L \tau
  \varphi(\alpha)}\,,
\end{equation}
where $\varphi$ is the dynamical counterpart to a free energy. It allows one to extend the language of phase transitions to dynamical systems \cite{Gaspard1998} and also yields the cumulants of $\lambda$ in the large $L$ limit since 

\begin{equation}
\mean{\lambda^{n}}_{c} = \frac{1}{(L
  \tau)^{n-1}} \left. \td{^{n} \varphi}{\alpha^n} \right|_{\alpha=0}.
\end{equation}

Enforcing the constraints with Lagrange multipliers, performing a perturbation expansion in $\alpha$ of the saddle-point equations and looking for stationary solutions, we get

\begin{equation}
  \varphi(\alpha) = -4 \pi^2 D(\rho_0)\, \alpha \left[1 - \frac{\alpha}{8} \frac{\kappa'(\rho_0)^2}{\kappa(\rho_0)^3} + \mathcal{O}(\alpha^2) \right]	\label{eqn:LDF}
\end{equation}
where $\kappa = \frac{2 D}{\sigma}$ is basically the compressibility. Note that the saddle-point equations yield $\dot \rho=-\partial_x \mathcal{J}$, showing that $\mathcal{J}$ can be seen as a particle current at the saddle-point level.

The analytical result~\eqref{eqn:LDF} is the first important result of our work. At this point, we notice that the mean value of the largest LE is negative and equal to $\varphi'(0) = - 4 \pi^{2} D(\rho_0)$, which corresponds to the largest LE of a diffusion equation with diffusity $D(\rho_0)$. All other LEs are thus also negative. This reflects the fact that diffusive dynamics tends to smooth out density profiles, hence eliminating perturbations rather than amplifying them. Taking $\alpha > 0$ in this case first detects less stable trajectories rather than chaotic ones. Equation~\eqref{eqn:LDF} can be extended, upon painful but systematic algebra, to higher order. For instance, we show here the series up to
the 5th order for the case $D=1$:

\begin{widetext}
	\begin{multline}
		\varphi(\alpha) = 
				- 4 \pi^2 \alpha 
				+ \frac{\pi^{2} \, \sigma'(\rho_{0})^{2}}{2 \, \sigma(\rho_{0})} \frac{\alpha^2}{2}
				- \frac{9 \pi^{2} \, \sigma'(\rho_0)^{4}}{2^{6} \, \sigma(\rho_0)^2} \frac{\alpha^3}{3!}\\
				+ \frac{3 \pi^{2} (55 \, \sigma'(\rho_0)^6 - 72 \, \sigma(\rho_0) \, \sigma'(\rho_0)^4 \, \sigma''(\rho_0) + 8 \, \sigma(\rho_0)^2 \, \sigma'(\rho_0)^2 \, \sigma''(\rho_0)^2 + 32 \, \sigma(\rho_0)^2 \, \sigma'(\rho_0)^3 \, \sigma'''(\rho_0))}{2^{10} \, \sigma(\rho_0)^3} \frac{\alpha^4}{4!}\\
				- \frac{15 \pi^2 (309 \, \sigma'(\rho_0)^8 - 512 \, \sigma(\rho_0) \, \sigma'(\rho_0)^6 \, \sigma''(\rho_0) + 256 \, \sigma(\rho_0)^2 \sigma'(\rho_0)^5 \, \sigma'''(\rho_0)} {2^{16} \, \sigma'(\rho_0)^4} \frac{\alpha^5}{5!}
				+ {\cal O}(\alpha^{6})
		\label{eqn:LDFD1}
	\end{multline}
\end{widetext}
\begin{floatequation}
\mbox{\textit{see eq.~\eqref{eqn:LDFD1}}}\,.
\end{floatequation}
We have not been able to infer a generic form of the coefficients from the first few contributions.

\begin{figure}[t]
	\onefigure[width=\linewidth]{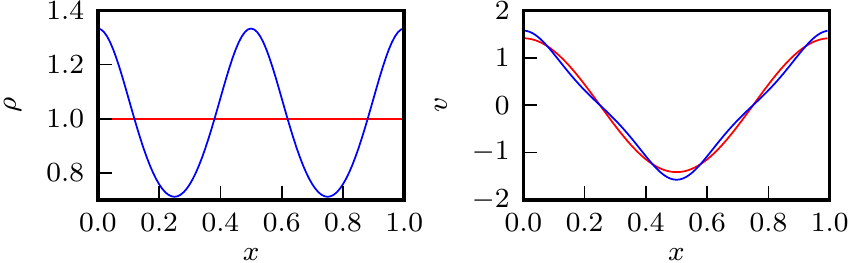}
	\caption{\label{fig:profiles} Energy profile $\rho(x)$ and tangent vector $v(x)$ of the KMP chain are shown for $\rho_0=1$ and $\alpha=0$ (typical case, in red) and $\alpha=0.4$ ($25\%$ increase of $\lambda$, in blue, with $\rho(x)$ strongly departing from uniformity). In the typical case ($\alpha = 0$), the dynamics amount to noiseless diffusion and the slowest decaying perturbation thus corresponds to a single Fourier mode, of smallest frequency. As $|\alpha|$ increases, nearby harmonics appear due to the non-linearity of the dynamics, causing the large-scale oscillations seen for $\alpha=0.4$.}
\end{figure}

Our approach can also be used to visualise how the system develops nontrivial structures to produce a Lyapunov exponent that deviates from its typical value, by calculating the density profiles $\rho(x)$ and tangent vector $v(x)$ -- both assumed stationary in time -- that extremise $S$ for a given value of $\alpha$.  Figure~\ref{fig:profiles} shows such realisations for $\alpha=0.4$, leading to a $25\%$ increase of the Lyapunov exponent~\footnote{For $\alpha=0.4$, the $\alpha^5$   term shown in eq. \eqref{eqn:LDFD1} amounts to $1\%$ of   $\varphi(\alpha)$.}. For this value, $\rho(x)$ is well approximated by a simple harmonic modulation but this would develop into more complex nonlinear shapes at larger $\alpha$.

\begin{figure}[t]
	\onefigure[width=\linewidth]{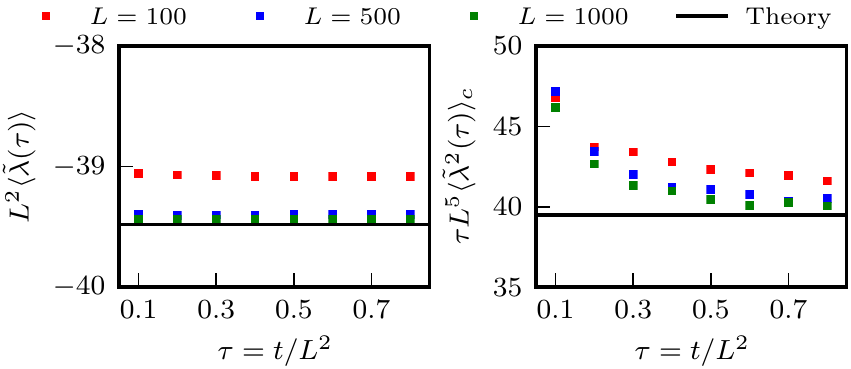}
	\caption{\label{fig:KMP_simu} Numerical evaluations of the mean and variance of ${\lambda}$ in the KMP model at different sizes and times, computed using $10^{6}$ samples for $L=100,\,500$ and $10^5$ samples for $L=1\,000$. Larger times are difficult to access because of rounding errors. The ${\cal O}(1/L)$ finite-size corrections to $\mean{\lambda}$ arise from the fluctuations around the saddle of the action \eqref{eqn:action}.}
\end{figure}

Since we have relied on a fluctuating hydrodynamic description, we would like to check whether the LE $\lambda$ calculated within this formalism is identical to the microscopic LE $\tilde{\lambda}$ of the original lattice model. We expect the ``discrete'' LE $\tilde{\lambda}(t)$ to be related to the ``fluctuating hydrodynamics'' LE $\lambda(\tau)$ by the relation $\tau\lambda(\tau)=\left.  t\tilde{\lambda}(t)\right|_{t=L^{2} \tau}$, i.e.~$\lambda=L^2\tilde\lambda$.  If this relation is correct, the cumulant of $\tilde\lambda$ should be given by $\mean{\tilde \lambda^{n}}_{c} = L^{1-3n}\tau^{1-n} \varphi^{(n)}(0)$ in the large-size and large-time limit. We have checked this numerically for the specific choice of the original KMP model. As can be seen in fig.~\ref{fig:KMP_simu}, the first two cumulants (mean and variance) are in good agreement with this prediction. The mean reaches its long-time limit for $\tau \sim {\cal O}(10^{-2})$ but the variance requires $\tau \sim {\cal O}(1)$. Our calculation of the LE in the hydrodynamic regime is thus fully consistent with the LE measured in the microscopic model. This shows that the MFT can indeed be extended to compute LEs of spatially extended diffusive systems.

In the second part of this Letter, we turn to models with discrete degrees of freedom and show an unexpected connection to damage spreading and annihilation processes. For the sake of concreteness, take a Symmetric Simple Exclusion Process (SSEP), in which particles perform a symmetric random walk with mutual exclusion. We consider a chain of size $L$, with unit hopping rate. In order to define the LE, we consider two copies $A$ and $B$ of this system, and we apply the same noise to both copies. Specifically, we assume that hops are triggered by the environment: when a site tries to expel a particle in one system, it will also expel a particle in the other (if there is one at this site). If $n_{i}^{A}$ and $n_{i}^{B}$ are respectively the occupation numbers at site $i$ in copy $A$ and in copy $B$, the local difference between the two copies is $u_{i} = n_{i}^{A} - n_{i}^{B}$, and we are interested in the evolution of its norm $\left| \mathbf{u} \right| = \sum_{i=1}^{L} \left| u_{i} \right|$. Here we chose the 1-norm because $p$-norms with $p>1$ are singular when studying macroscopic effects as $u_i=0,\pm1$ and thus $\left| u_{i} \right|^{p} = \left| u_{i} \right|$.

Readers will now realise that the calculation of the LE is closely linked to the issue of damage spreading~\cite{Derrida1987, Glotzer1991, Vojta1997}. There, one studies the propagation of a spatial defect, i.e.~a small difference between two nearly identical copies of the same system, over time and asks whether this defect spreads or recedes. With the LE, one further looks at the rate at which the defect vanishes or completely changes the subsequent evolution of the system. 
\begin{figure}
	\centering
	\begin{tabular}[c]{|c|c|}
		\hline
		Coupled-SSEPs dynamics  & 		Effective dynamics for $u_{i}$\\
   		\hline
      			
 	  	% Première ligne	
		\includegraphics[width=0.3\linewidth]{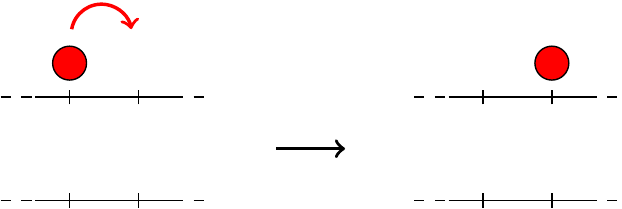}
			
		&

		\includegraphics[width=0.3\linewidth]{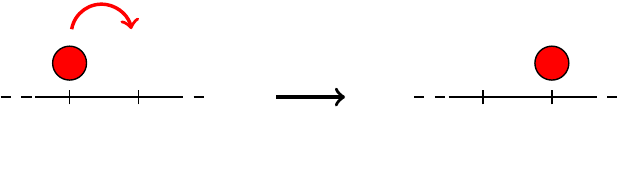}
					
		\\
			
		\hline
				
		% Deuxième ligne
		\includegraphics[width=0.3\linewidth]{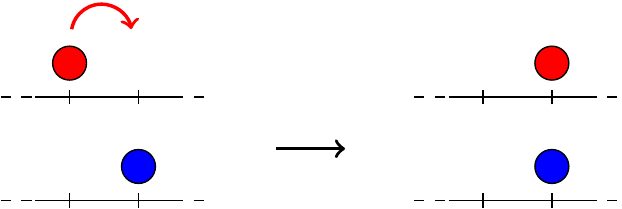}
		
		&

		\includegraphics[width=0.3\linewidth]{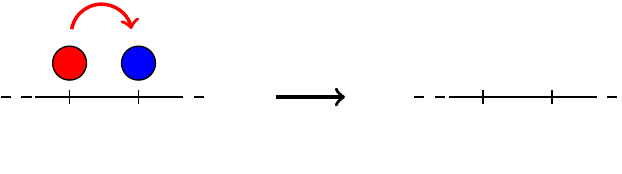}
				
		\\
				
		\hline
				
		% Troisème ligne
		\includegraphics[width=0.3\linewidth]{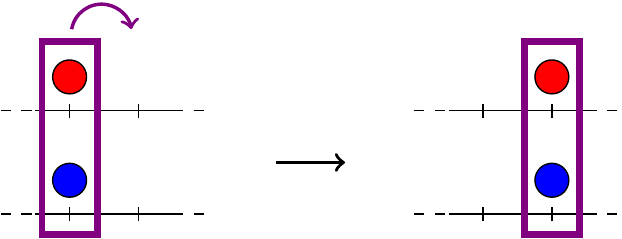}
			
		&

		\includegraphics[width=0.3\linewidth]{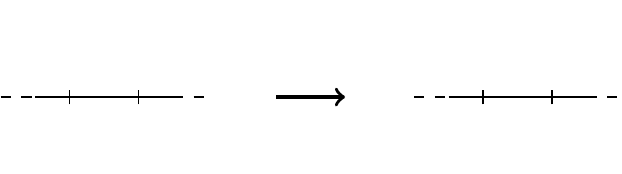}
				
		\\
				
		\hline
				
		% Quatrième ligne
		\includegraphics[width=0.3\linewidth]{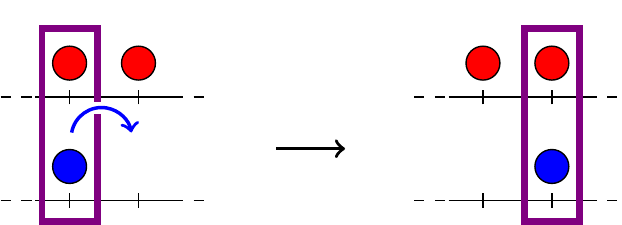}
				
		&

		\includegraphics[width=0.3\linewidth]{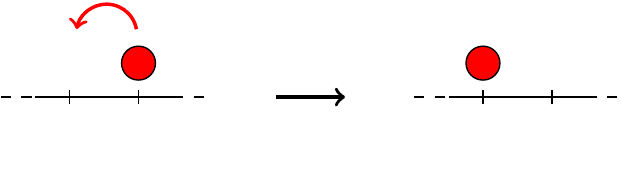}
				
		\\
				
		\hline
	
	\end{tabular}
	
	\caption{Effective dynamics for $\mathbf{u}$ in the two coupled SSEPs. We look at all the cases when particles at a given site try to hop to the right. In the left column, the top chain is copy $A$ and the bottom one copy $B$. In the right column, a red particle represents $u_{i} = +1$ and a blue one $u_{i} = -1$. This is summarised in fig. \ref{fig:mapping}}
	\label{fig:detail_mapping}
\end{figure}

\begin{figure}[t]
	\onefigure[width=\linewidth]{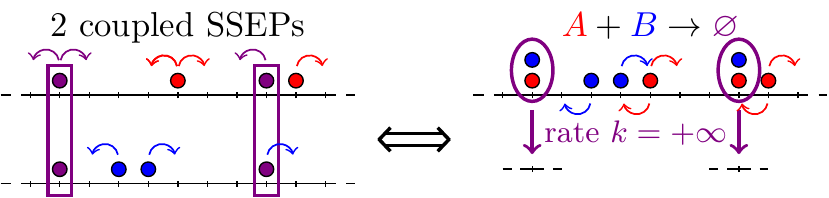}
	\caption{Mapping between two coupled SSEPs and the pair annihilation reaction-diffusion process ${A + B \rightarrow \varnothing}$. The violet particles in the two coupled SSEPs, occupying the same sites on each lattice, do not affect the subsequent evolution of $u_i=n_i^A-n_i^B$. For instance, if a fluctuation attempts to make the rightmost violet particles in the SSEPs hop to the right, only the bottom one does so. This is then equivalent to a hop to the left of the rightmost red particle in the reaction-diffusion process, as far as $u_i$ is concerned.}
	\label{fig:mapping} 
\end{figure}

\begin{figure}[!t]
	\onefigure[width=\linewidth]{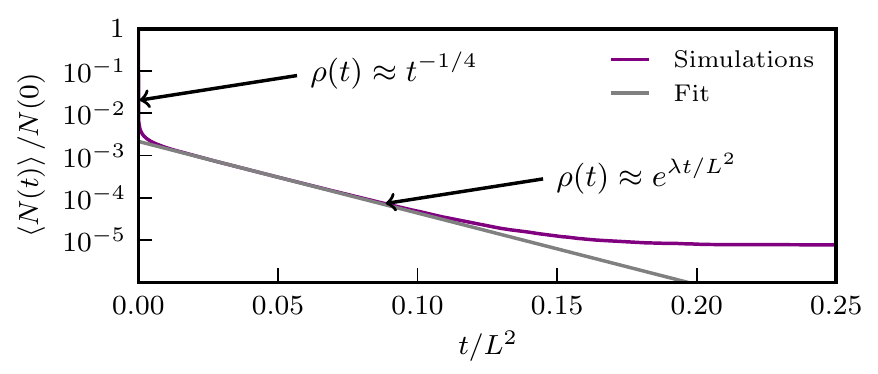}
	\caption{\label{fig:AB0}Simulations of $A + B \rightarrow \varnothing$ performed using the RRC algorithm~\cite{Avraham1988}. The system size is $2^{20}$, the initial density of each species of particle is $1/8$, and the average is performed over the steady-state distribution extracted from $1\,000$ runs, conditioned on not being absorbed. After a power-law decay, with an exponent $-1/4$ predicted for the infinite size limit, we see the emergence of an exponential decay due to the finite size of the system. The decay rate measured numerically is $\lambda \approx -39$ in units of $\tau$, which is within few percent of $\lambda=-4\pi^2$ predicted using our mapping to two coupled SSEPs.}
\end{figure}

The dynamics of $u_i = n_{i}^{A} - n_{i}^{B}$ in the two coupled SSEPs can be mapped onto the same quantity in the two-species pair annihilation reaction-diffusion process ${A + B \rightarrow   \varnothing}$~\cite{Zeldovich1978, Toussaint1983}.  In this two-species models, the $A$ particles perform a symmetric random walk with mutual exclusion, the $B$ particles do the same, and when $A$ and $B$ particles meet at the same site, they immediately annihilate. The mapping arises from the fact that when a site is occupied in both SSEPs, removing the particles in both systems will not affect the subsequent dynamics of $u_i$ (see fig.~\ref{fig:detail_mapping} and fig.~\ref{fig:mapping}).  The asymptotics of the ${A + B \rightarrow \varnothing}$ process are fully understood for infinite system size \cite{Bramson1988}, but there are no exact results in finite size for averages let alone fluctuations. Since the SSEP can be described by fluctuating hydrodynamics with $D(\rho) = 1$ and $\sigma(\rho) = 2 \rho (1 - \rho)$, we know from eq.~\eqref{eqn:LDF} that $\left| \mathbf{u} (t)\right| \approx e^{\tilde \lambda t}$ with $\langle\tilde \lambda\rangle=-\frac{4\pi^2}{L^2}$. Hence, thanks to our mapping, we can predict that in the large-size and large-time limit, the total number $N(t)$ of particles in the ${A+B\rightarrow\varnothing}$ process scales as

\begin{equation}
  {\mean{N(t)} \approx e^{- 4 \pi^{2} {t}/{L^{2}}}}.
\end{equation}
This regime, which was out of reach of previous numerical studies~\cite{Simon1995, Lee2000}, is in perfect agreement with our simulations (see fig.~\ref{fig:AB0}). Note that we also have predictions for the fluctuations of $N(t)$, from eq.~\eqref{eqn:LDF}, but confirming these numerically is difficult since the absorbing (empty) state is reached too quickly.

In this letter, we have shown how a generalisation of the MFT can be used to compute analytically the fluctuations of the largest LE of spatially extended systems described by diffusive fluctuating hydrodynamics. The relevance of our approach has been confirmed by direct comparison to a microscopic model. Interestingly, the mapping of the SSEP to ${A + B \rightarrow \varnothing}$ suggests a generic correspondence between damage spreading/LE determination for systems with discrete degrees of freedom and reaction-diffusion processes with absorbing states. This would be an interesting direction to pursue. But perhaps as challenging would be to exploit similar techniques to study suspensions of interacting colloids, which could open the way to identifying slow modes in glass formers. A concrete starting point would be the stochastic evolution equation established by Dean~\cite{Dean1996}, for which the strategy deployed in this work would apply, though alternative approximation schemes would have to be adopted. Leaving the realm of glasses for that of dynamical systems, it would be interesting to apply our approach to the recently derived fluctuating hydrodynamics of the FPU chain~\cite{Mendl2013,Das2014}, for which large deviations of the largest LE are associated with the emergence of breathers and solitons~\cite{Tailleur2007,Laffargue2013}.

We warmly acknowledge early discussions with Peter Grassberger and Henk van Beijeren. JT thanks the Galileo Galilei Institute for Theoretical Physics for the hospitality and the INFN for partial support during the completion of this work. FvW acknowledges the support of the Pitzer Center of the UC Berkeley's Department of Chemistry.

%\bibliography{/home/tanguy/Documents/JabRef/Tanguy}
%\bibliography{./Tanguy.bib}
\bibliography{./Chaos_letter.bib}

\begin{thebibliography}{10}
\expandafter\ifx\csname url\endcsname\relax\def\url#1{\texttt{#1}}\fi

\bibitem{Dorfman1999}
\Name{Dorfman J.~R.} \Book{An Introduction to Chaos in Nonequilibrium
  Statistical Mechanics} Vol.~14 of \emph{Cambridge Lecture Notes in Physics}
  (Cambridge University Press, Cambridge) 1999.
\newline\url{http://books.google.fr/books?id=Mqlff1KbjMQC}

\bibitem{Penrose1979}
\Name{Penrose O.} \REVIEW{Rep. Prog. Phys.}{42}{1979}{1937}.

\bibitem{Gaspard1998a}
\Name{Gaspard P., Briggs M.~E., Francis M.~K., Sengers J.~V., Gammon R.~W.,
  Dorfman J.~R. \and Calabrese R.~V.} \REVIEW{Nature}{394}{1998}{865}.

\bibitem{Dettmann1999}
\Name{Dettmann C.~P., Cohen E. G.~D. \and Van~Beijeren H.}
  \REVIEW{Nature}{401}{1999}{875}.

\bibitem{Grassberger1999}
\Name{Grassberger P. \and Schreiber T.} \REVIEW{Nature}{401}{1999}{875}.

\bibitem{Evans1993}
\Name{Evans D.~J., Cohen E. G.~D. \and Morriss G.~P.} \REVIEW{Phys. Rev.
  Lett.}{71}{1993}{2401}.

\bibitem{Gallavotti1995}
\Name{Gallavotti G. \and Cohen E. G.~D.} \REVIEW{Phys. Rev.
  Lett.}{74}{1995}{2694}.

\bibitem{Benzi1984}
\Name{Benzi R., Paladin G., Parisi G. \and Vulpiani A.} \REVIEW{J. Phys.
  A}{17}{1984}{3521}.

\bibitem{Arnold1986}
\Name{Arnold L., Kliemann W. \and Oeljeklaus E.} \Book{Lyapunov exponents of
  linear stochastic systems} in \Book{Lyapunov Exponents}, edited by
  \Name{Arnold L. \and Wihstutz V.} Vol. 1186 of \emph{Lecture Notes in
  Mathematics} (Springer Berlin Heidelberg) 1986 pp. 85--125.

\bibitem{Graham1988}
\Name{Graham R.} \REVIEW{Europhys. Lett.}{5}{1988}{101}.

\bibitem{Paladin1995}
\Name{Paladin G., Serva M. \and Vulpiani A.} \REVIEW{Phys. Rev.
  Lett.}{74}{1995}{66}.

\bibitem{Derrida2004}
\Name{Derrida B., Douçot B. \and Roche P.~E.} \REVIEW{J. Stat.
  Phys.}{115}{2004}{717}.

\bibitem{Bodineau2004}
\Name{Bodineau T. \and Derrida B.} \REVIEW{Phys. Rev. Lett.}{92}{2004}{180601}.

\bibitem{Bertini2005}
\Name{Bertini L., De~Sole A., Gabrielli D., Jona-Lasinio G. \and Landim C.}
  \REVIEW{Phys. Rev. Lett.}{94}{2005}{030601}.

\bibitem{Bertini2005a}
\Name{Bertini L., Gabrielli D. \and Lebowitz J.~L.} \REVIEW{J. Stat.
  Phys.}{121}{2005}{843}.

\bibitem{Imparato2009}
\Name{Imparato A., Lecomte V. \and van Wijland F.} \REVIEW{Phys. Rev.
  E}{80}{2009}{011131}.

\bibitem{Brunet2010}
\Name{Brunet E., Derrida B. \and Gerschenfeld A.} \REVIEW{Europhys.
  Lett.}{90}{2010}{20004}.

\bibitem{Lecomte2010}
\Name{Lecomte V., Imparato A. \and van Wijland F.} \REVIEW{Progr. Theoret.
  Phys. Suppl.}{184}{2010}{276}.

\bibitem{Hurtado2011}
\Name{Hurtado P.~I. \and Garrido P.~L.} \REVIEW{Phys. Rev.
  Lett.}{107}{2011}{180601}.

\bibitem{Meerson2014}
\Name{Meerson B. \and Sasorov P.~V.} \REVIEW{Phys. Rev. E}{89}{2014}{010101}.

\bibitem{Garrahan2007}
\Name{Garrahan J.~P., Jack R.~L., Lecomte V., Pitard E., van Duijvendijk K.
  \and van Wijland F.} \REVIEW{Phys. Rev. Lett.}{98}{2007}{195702}.

\bibitem{Turci2011}
\Name{Turci F. \and Pitard E.} \REVIEW{Europhys. Lett.}{94}{2011}{10003}.

\bibitem{Lecomte2012}
\Name{Lecomte V., Garrahan J.~P. \and van Wijland F.} \REVIEW{J. Phys.
  A}{45}{2012}{175001}.

\bibitem{Speck2012}
\Name{Speck T. \and Chandler D.} \REVIEW{J. Chem. Phys.}{136}{2012}{184509}.

\bibitem{Merolle2005}
\Name{Merolle M., Garrahan J.~P. \and Chandler D.} \REVIEW{Proc. Natl. Acad.
  Sci. U.S.A.}{102}{2005}{10837}.

\bibitem{Hedges2009}
\Name{Hedges L.~O., Jack R.~L., Garrahan J.~P. \and Chandler D.}
  \REVIEW{Science}{323}{2009}{1309}.

\bibitem{Pitard2011}
\Name{Pitard E., Lecomte V. \and van Wijland F.} \REVIEW{Europhys.
  Lett.}{96}{2011}{56002}.

\bibitem{Fullerton2013}
\Name{Fullerton C.~J. \and Jack R.~L.} \REVIEW{J. Chem.
  Phys.}{138}{2013}{224506}.

\bibitem{Ruelle1978}
\Name{Ruelle D.} \Book{Thermodynamic Formalism: The Mathematical Structure of
  Equilibrium Statistical Mechanics} (Addison-Wesley) 1978.

\bibitem{Gaspard1998}
\Name{Gaspard P.} \Book{Chaos, Scattering and Statistical Mechanics} (Cambridge
  University Press) 1998.

\bibitem{Bohr1987}
\Name{Bohr T. \and Rand D.} \REVIEW{Physica D}{25}{1987}{387}.

\bibitem{Grassberger1988}
\Name{Grassberger P., Badii R. \and Politi A.} \REVIEW{J. Stat.
  Phys.}{51}{1988}{135}.

\bibitem{Beck1993}
\Name{Beck C. \and Schögl F.} \Book{Thermodynamics of Chaotic Systems: An
  Introduction} (Cambridge University Press) 1993.

\bibitem{Appert1997}
\Name{Appert C., van Beijeren H., Ernst M. \and Dorfman J.} \REVIEW{J. Stat.
  Phys.}{87}{1997}{1253}.

\bibitem{Kuptsov2011}
\Name{Kuptsov P.~V. \and Politi A.} \REVIEW{Phys. Rev.
  Lett.}{107}{2011}{114101}.

\bibitem{Pazo2013}
\Name{Pazó D., López J.~M. \and Politi A.} \REVIEW{Phys. Rev.
  E}{87}{2013}{062909}.

\bibitem{Tailleur2007}
\Name{Tailleur J. \and Kurchan J.} \REVIEW{Nat. Phys.}{3}{2007}{203}.

\bibitem{Laffargue2013}
\Name{Laffargue T., Nguyen Thu~Lam K.-D., Kurchan J. \and Tailleur J.}
  \REVIEW{J. Phys. A}{46}{2013}{254002}.

\bibitem{Takeuchi2013}
\Name{Takeuchi K.~A. \and Chaté H.} \REVIEW{J. Phys. A: Math.
  Theor.}{46}{2013}{254007}.

\bibitem{Yang2009}
\Name{Yang H.-l., Takeuchi K.~A., Ginelli F., Chaté H. \and Radons G.}
  \REVIEW{Phys. Rev. Lett.}{102}{2009}{074102}.

\bibitem{Touchette2006}
\Name{Touchette H. \and Beck C.} \REVIEW{J. Stat. Phys.}{125}{2006}{455}.

\bibitem{Caglioti1996}
\Name{Caglioti E. \and Loreto V.} \REVIEW{Phys. Rev. E}{53}{1996}{2953}.

\bibitem{Kipnis1982}
\Name{Kipnis C., Marchioro C. \and Presutti E.} \REVIEW{J. Stat.
  Phys.}{27}{1982}{65}.

\bibitem{Spohn1983}
\Name{Spohn H.} \REVIEW{J. Phys. A}{16}{1983}{4275}.

\bibitem{Bertini2001}
\Name{Bertini L., De~Sole A., Gabrielli D., Jona-Lasinio G. \and Landim C.}
  \REVIEW{Phys. Rev. Lett.}{87}{2001}{040601}.

\bibitem{Bertini2002}
\Name{Bertini L., De~Sole A., Gabrielli D., Jona-Lasinio G. \and Landim C.}
  \REVIEW{J. Stat. Phys.}{107}{2002}{635}.

\bibitem{Tailleur2008}
\Name{Tailleur J., Kurchan J. \and Lecomte V.} \REVIEW{J. Phys. A: Math.
  Theor.}{41}{2008}{505001}.

\bibitem{Derrida2009a}
\Name{Derrida B. \and Gerschenfeld A.} \REVIEW{J. Stat. Phys.}{137}{2009}{978}.

\bibitem{Bouchet2014}
\Name{Bouchet F., Laurie J. \and Zaboronski O.} \REVIEW{J. Stat.
  Phys.}{156}{2014}{1066}.

\bibitem{Janssen1976}
\Name{Janssen H.-K.} \REVIEW{Z. Phys. B}{23}{1976}{377}.

\bibitem{DeDominicis1976}
\Name{De~Dominicis C.} \REVIEW{J. Phys. Colloques}{37}{1976}{C1}.

\bibitem{Derrida1987}
\Name{Derrida B.} \REVIEW{J. Phys. A}{20}{1987}{L721}.

\bibitem{Glotzer1991}
\Name{Glotzer S.~C. \and Jan N.} \REVIEW{Physica A}{173}{1991}{325}.

\bibitem{Vojta1997}
\Name{Vojta T.} \REVIEW{J. Phys. A}{30}{1997}{L7}.

\bibitem{Zeldovich1978}
\Name{Zeldovich Y.~B. \and Khlopov M.~Y.} \REVIEW{Physics Letters
  B}{79}{1978}{239}.

\bibitem{Toussaint1983}
\Name{Toussaint D. \and Wilczek F.} \REVIEW{J. Chem. Phys}{78}{1983}{2642}.

\bibitem{Bramson1988}
\Name{Bramson M. \and Lebowitz J.~L.} \REVIEW{Phys. Rev.
  Lett.}{61}{1988}{2397}.

\bibitem{Simon1995}
\Name{Simon H.} \REVIEW{J. Phys. A}{28}{1995}{6585}.

\bibitem{Lee2000}
\Name{Lee J.~W.} \REVIEW{Phys. Rev. E}{62}{2000}{2959}.

\bibitem{Avraham1988}
\Name{ben Avraham D.} \REVIEW{J. Chem. Phys}{88}{1988}{941}.

\bibitem{Dean1996}
\Name{Dean D.} \REVIEW{J. Phys. A}{29}{1996}{L613}.

\bibitem{Mendl2013}
\Name{Mendl C.~B. \and Spohn H.} \REVIEW{Phys. Rev. Lett.}{111}{2013}{230601}.

\bibitem{Das2014}
\Name{Das S.~G., Dhar A., Saito K., Mendl C.~B. \and Spohn H.} \REVIEW{Phys.
  Rev. E}{90}{2014}{012124}.

\end{thebibliography}

\end{document}